\documentclass[12pt]{article}
\usepackage{float}
\usepackage{authblk}
\usepackage{amssymb,amsmath}
\usepackage{graphicx}
\usepackage{subfigure}
\usepackage{caption}
\usepackage{subcaption}
\usepackage{cite}
\usepackage{xcolor}
\usepackage{setspace}
\usepackage{appendix}
\usepackage{makecell}
\usepackage{array}
\usepackage{tabularx}
\newcolumntype{M}[1]{>{\centering\arraybackslash}m{#1}}
\usepackage[colorlinks=true,urlcolor=blue,citecolor=blue,linkcolor=red,linktocpage=true,pdfproducer=medialab]{hyperref}
\usepackage[a4paper,width=17cm,height=25cm]{geometry}
\usepackage{morefloats}
\usepackage{siunitx}
\makeatletter
\renewcommand{\@dotsep}{10000}
\makeatother
\allowdisplaybreaks[4]

\begin{document}
	\title{QCD sum rule analysis of hidden strange $1^{++}$ tetraquark masses and radial excitations}
	\author[1,2]{Zhuo-Ran Huang}
	\author[2,3]{Wei Chen}
	\author[4]{Jason Ho}
	\author[1]{Lei-Hua Liu}
	\affil[1]{College of Physics and Mechanical and Electrical Engineering, Jishou University, Jishou 416000, China}
	\affil[2]{School of Physics, Sun Yat-Sen University, Guangzhou 510275, China}
	\affil[3]{Southern Center for Nuclear-Science Theory (SCNT), Institute of Modern Physics, Chinese Academy of Sciences, Huizhou 516000, Guangdong Province, China}
	\affil[4]{Department of Physics, Dordt University, Sioux Center, Iowa, 51250, USA}
	\date{\today}
		\maketitle
		\begin{abstract}
			We revisit the light tetraquark states with quantum numbers $J^{PC}=1^{++}$ using QCD sum rules, focusing on a complete set of derivative-free diquark-antidiquark interpolating currents. By calculating the operator product expansion up to dimension-eight condensates, we extract the ground-state mass of the hidden-strange $us\bar{u}\bar{s}$ tetraquark from both Laplace sum rules (LSR) and finite-energy sum rules (FESR). Our combined analysis yields $M_{us\bar{u}\bar{s}} = 1.45\pm0.11$~GeV, which agrees well with the mass of the $a_1(1420)$ resonance and supports its tetraquark interpretation. Furthermore, we perform Gaussian sum rule (GSR) analyses to probe radial excitations, adopting a two-resonance narrow-width model. The GSR fit reveals a heavier state with mass $m_2 = 1.86\pm0.12$~GeV and a relative coupling $r = 0.15\pm0.03$ for the lighter state, indicating that the $a_1(1930)$ is a promising candidate for a compact $1^{++}$ tetraquark. We also discuss the dominant decay modes of these tetraquark candidates, emphasizing hidden-strange channels such as $K^*K$ and $f_0(980)\pi$, which can be tested in future experiments.
		\end{abstract}
\section{Introduction}
Within the traditional quark model, mesons are understood as \(q\bar{q}\) bound states, with quantum numbers constrained by \(P = (-1)^{L+1}\) and \(C = (-1)^{L+S}\) for neutral systems. The axial-vector channel with \(J^{PC}=1^{++}\) is naturally realized, for instance, by the \(L=1\), \(S=1\) configuration, and is populated by well-known resonances such as \(a_1(1260)\) and \(f_1(1285)\). However, the experimental spectrum of light axial-vector mesons has become remarkably rich: the COMPASS Collaboration observed a narrow \(1^{++}\) signal in the \(f_0(980)\pi\) channel, identified as \(a_1(1420)\) with mass \(1414^{+15}_{-13}\)~MeV and width \(153^{+8}_{-23}\)~MeV~\cite{COMPASS:2015kdx}. This state joins a growing family of \(a_1\) resonances---\(a_1(1260)\), \(a_1(1420)\), \(a_1(1640)\), \(a_1(1930)\), \(a_1(2095)\), \(a_1(2270)\)~\cite{ParticleDataGroup:2024cfk}---whose overpopulation suggests the presence of exotic structures beyond the simple quark-antiquark picture. In particular, the strong coupling of \(a_1(1420)\) to \(f_0(980)\pi\), where \(f_0(980)\) is known to contain a significant \(\bar{s}s\) component, hints at a possible tetraquark interpretation.

The QCD sum rule approach~\cite{Shifman:1978bx,Reinders:1984sr} provides a powerful, nonperturbative tool to relate hadronic properties to the fundamental parameters of QCD. Over the years, light tetraquark systems with various quantum numbers have been studied using LSR and FESR~\cite{Huang:2016rro,Fu:2018ngx,Sundu:2017xct,Chen:2008qw}. For the \(1^{++}\) channel, a systematic classification of tetraquark interpolating currents was carried out by Chen et al.~\cite{Chen:2015fwa,Chen:2013jra}, who divided the currents into antisymmetric (A), symmetric (S) and mixed (M) types according to the flavor symmetries of the constituent diquarks and antidiquarks. Using LSR with currents of type~M that contain one \(s\bar{s}\) pair (\(qs\bar{q}\bar{s}\)), they obtained a mass \(1.44\pm0.08\)~GeV, in excellent agreement with the \(a_1(1420)\) resonance, and proposed that its isoscalar partner is the well-established \(f_1(1420)\).

In parallel with the tetraquark interpretation, a number of alternative dynamical explanations for the \(a_1(1420)\) structure have been proposed. It was suggested that the narrow enhancement could originate from a triangle singularity (TS) mechanism, wherein the decay \(a_1(1260)\to K^*\bar{K}\) followed by \(K^*\to K\pi\) and \(K\bar{K}\to f_0(980)\) produces a kinematic peak without a genuine resonance pole~\cite{Mikhasenko2015,Aceti2016,Guo2020}. Subsequent works have refined this picture by incorporating final-state interactions and coupled-channel effects within unitary three-body frameworks~\cite{Sakthivasan2024,Sakthivasan:2026bph}. More recently, a virtual-state interpretation has also been put forward~\cite{Yan2025}. These studies highlight that the observed lineshape can be mimicked by non-resonant dynamics, challenging the necessity of a compact tetraquark state. However, the QCD sum-rule approach—which directly probes the intrinsic quark-gluon structure—offers complementary information that can distinguish between a genuine multiquark bound state and a purely kinematic effect.

Nevertheless, early QCD sum-rule analyses of the light $1^{++}$ tetraquark states included the contributions of high-dimensional condensates ($d>8$) arising from single-propagator contributions, which are incomplete. The violation of factorization of higher dimensional condensates was also not considered. Moreover, most previous work focused exclusively on LSR and did not systematically explore alternative sum-rule techniques such as FESR or GSR~\cite{Ho:2019org,Ho:2018cat} for radial excitations.

In this paper we revisit the light \(1^{++}\) tetraquark sector by constructing a complete set of \textit{derivative-free} interpolating currents. These low-dimensional currents are expected to yield a better-behaved OPE series and therefore more reliable mass predictions. We explicitly list eight independent currents (denoted \(J_{1\mu}\) through \(J_{8\mu}\)) that carry a single Lorentz index and are built from diquark-antidiquark pairs without any covariant derivatives. For each current we compute the OPE up to dimension-eight condensates and extract the corresponding spectral densities. Using the spectral densities, we perform LSR and FESR to obtain the mass of the ground state of the \(1^{++}\) tetraquarks, and GSR to make predictions for the first radial excitation.

A brief discussion of the dominant decay modes of these tetraquark candidates is also provided, complementing the analysis of Ref.~\cite{Chen:2015fwa}.

This paper is organized as follows. Section~2 recalls the LSR and FESR formalism. In Sec.~3 we construct the complete set of derivative-free tetraquark currents with \(J^{PC}=1^{++}\) and outlines the OPE calculation. Numerical results from LSR and FESR are presented in Sec.~4, while Sec.~5 is devoted to the GSR study of radial excitations. Section~6 discusses possible decay patterns, and Sec.~8 concludes.
\section{Laplace Sum Rules and Finite Energy Sum Rules}
The QCD sum-rule formalism \cite{Shifman:1978bx,Reinders:1984sr,Colangelo:2000dp} provides a bridge between the hadronic world and the underlying quark–gluon dynamics through the analytic properties of correlation functions. For a local current $j_\mu(x)$ carrying the quantum numbers of an axial-vector state, we consider the two-point correlation function
\begin{equation}
	\Pi_{\mu\nu}(q^2) = i\int d^4x\, e^{iqx}\langle 0|T[\,j_\mu(x)\,j_\nu^\dagger(0)\,]|0\rangle 
	= (q_\mu q_\nu - q^2 g_{\mu\nu})\,\Pi_A(q^2) + q_\mu q_\nu\,\Pi_P(q^2),
	\label{eq:corr}
\end{equation}
where the invariant functions $\Pi_A(q^2)$ and $\Pi_P(q^2)$ receive contributions from intermediate states with spin-parity $1^+$ and $0^-$, respectively. Our interest lies solely in the axial-vector channel $\Pi_A(q^2)$, which contains the $1^{++}$ tetraquark signal.

The invariant amplitude $\Pi_A(q^2)$ satisfies a dispersion relation with one subtraction, which in the deep Euclidean region ($Q^2 = -q^2 > 0$) can be written as
\begin{equation}
	\Pi_A(Q^2) = \frac{1}{\pi}\int_0^\infty \frac{\operatorname{Im}\Pi_A(s)}{s+Q^2}\,ds + \text{subtractions},
	\label{eq:dispersion}
\end{equation}
where the spectral function $\operatorname{Im}\Pi_A(s)$ encodes the hadronic spectrum. In the QCD sum-rule approach, $\Pi_A(Q^2)$ is evaluated in the operator product expansion (OPE) as a power series in $1/Q^2$:
\begin{equation}
	\Pi_A^{\mathrm{OPE}}(Q^2) = \sum_{n=0}^{\infty} \frac{\mathcal{C}_n(Q^2,\mu)}{(Q^2)^n} \langle \mathcal{O}_n(\mu) \rangle,
	\label{eq:ope}
\end{equation}
where $\mathcal{C}_n$ are perturbative Wilson coefficients and $\langle \mathcal{O}_n \rangle$ are vacuum condensates of increasing mass dimension. In this work we include condensates up to dimension 8, which is sufficient to achieve a convergent OPE for the light tetraquark currents under study.

To suppress the contributions from higher-dimensional condensates and to enhance the lowest-lying resonance, one applies the Borel transformation, defined by
\begin{equation}
	\mathcal{B}_{Q^2} = \lim_{\substack{Q^2,n\to\infty\\ Q^2/n\equiv \tau}} \frac{(Q^2)^n}{(n-1)!}\left(-\frac{d}{dQ^2}\right)^n,
	\label{eq:borel}
\end{equation}
which transforms the OPE series (with its polynomial subtractions) into an exponentially weighted integral over the spectral function. Acting on $\Pi_A(Q^2)$ and using the dispersion relation, we obtain the LSR moment
\begin{equation}
	M(\tau, s_0) = \int_0^{s_0} ds\, e^{-s\tau} \rho(s),
	\label{eq:lsr}
\end{equation}
where $\tau$ is the Borel parameter, $s_0$ the continuum threshold, and $\rho(s) \equiv \frac{1}{\pi}\operatorname{Im}\Pi_A(s)$ is the hadronic spectral density. The upper limit $s_0$ in the integral reflects the assumption that for $s>s_0$ the spectral density is well approximated by its QCD continuum expression, which is obtained from the OPE and cancels the corresponding perturbative contributions.

Adopting a ``single narrow resonance'' plus continuum parametrization,
\begin{equation}
	\rho(s) \simeq f_H^2 \,\delta(s-m_H^2) + \theta(s-s_0)\,\rho^{\mathrm{OPE}}(s),
	\label{eq:rho}
\end{equation}
the mass of the ground state can then be extracted from the logarithmic derivative of the resonance contribution:
\begin{equation}
	R(\tau, s_0) = -\frac{d}{d\tau}\ln M(\tau,s_0) = M_H^2,
	\label{eq:logder}
\end{equation}
provided that the continuum contribution is properly subtracted. In practice, we evaluate $R$ using the full OPE expression of $M(\tau,s_0)$ and search for stability regions in the $(\tau,s_0)$ plane where the dependence on both parameters is mild. 

An alternative and complementary method is the FESR, which is obtained from the analyticity of $\Pi_A(q^2)$ without performing the Borel transformation. Starting from the contour integral
\begin{equation}
	\frac{1}{2\pi i}\oint_{|q^2|=s_0} dq^2\, (q^2)^n \Pi_A(q^2) = 0
	\label{eq:contour}
\end{equation}
for sufficiently large $n$ (or by integrating the dispersion relation with a suitable weight), one derives the FESR moment
\begin{equation}
	W(n, s_0) = \int_0^{s_0} ds\, s^n \rho(s) = \frac{1}{\pi}\int_0^{s_0} ds\, s^n \operatorname{Im}\Pi_A^{\mathrm{OPE}}(s),
	\label{eq:fesr}
\end{equation}
which, in the OPE side, is expressed in terms of the spectral density computed from the OPE. The ratio of consecutive moments gives the squared mass:
\begin{equation}
	M_H^2(n, s_0) = \frac{W(n+1, s_0)}{W(n, s_0)}.
	\label{eq:fesr_mass}
\end{equation}
In practice we employ the zeroth moment ($n=0$), which benefits from an improved convergence of the expansion and suppresses the contributions of high-dimensional condensates. The FESR approach is particularly useful when the LSR displays poor OPE behaviour due to rapid fall-off of the exponential weight, a situation that can occur for multiquark states. By comparing the results obtained from both LSR and FESR, we can assess the systematic uncertainties related to the sum-rule method and the modelling of the spectral function.
\section{Tetraquark Interpolating Currents and QCD Expressions}
To study the $1^{++}$ tetraquark states we construct diquark–antidiquark interpolating currents without any covariant derivatives. The building blocks are the six independent diquark operators in Dirac space:
$q_a^T C q_b$, $q_a^T C\gamma_5 q_b$, $q_a^T C\gamma_\mu q_b$, $q_a^T C\gamma_\mu\gamma_5 q_b$, $q_a^T C\sigma_{\mu\nu} q_b$, $q_a^T C\sigma_{\mu\nu}\gamma_5 q_b$,
where $C$ is the charge-conjugation matrix and $a,b$ are color indices. By coupling a diquark with an antidiquark and (anti)symmetrizing appropriately, we obtain the following eight independent currents that carry one Lorentz index~\cite{Chen:2010ze}:
\begin{align}
	\begin{aligned}
		J_{1\mu} &= u_a^T C s_b \, (\bar{u}_a \gamma_\mu \gamma_5 C \bar{s}_b^T + \bar{u}_b \gamma_\mu \gamma_5 C \bar{s}_a^T) + u_a^T C \gamma_\mu \gamma_5 s_b \, (\bar{u}_a C \bar{s}_b^T + \bar{u}_b C \bar{s}_a^T), \\[4pt]
		J_{2\mu} &= u_a^T C s_b \, (\bar{u}_a \gamma_\mu \gamma_5 C \bar{s}_b^T - \bar{u}_b \gamma_\mu \gamma_5 C \bar{s}_a^T) + u_a^T C \gamma_\mu \gamma_5 s_b \, (\bar{u}_a C \bar{s}_b^T - \bar{u}_b C \bar{s}_a^T), \\[4pt]
		J_{3\mu} &= u_a^T C \gamma_5 s_b \, (\bar{u}_a \gamma_\mu C \bar{s}_b^T + \bar{u}_b \gamma_\mu C \bar{s}_a^T) + u_a^T C \gamma_\mu s_b \, (\bar{u}_a \gamma_5 C \bar{s}_b^T + \bar{u}_b \gamma_5 C \bar{s}_a^T), \\[4pt]
		J_{4\mu} &= u_a^T C \gamma_5 s_b \, (\bar{u}_a \gamma_\mu C \bar{s}_b^T - \bar{u}_b \gamma_\mu C \bar{s}_a^T) + u_a^T C \gamma_\mu s_b \, (\bar{u}_a \gamma_5 C \bar{s}_b^T - \bar{u}_b \gamma_5 C \bar{s}_a^T), \\[4pt]
		J_{5\mu} &= u_a^T C \gamma^\nu s_b \, (\bar{u}_a \sigma_{\mu\nu}\gamma_5 C \bar{s}_b^T + \bar{u}_b \sigma_{\mu\nu}\gamma_5 C \bar{s}_a^T) + u_a^T C \sigma_{\mu\nu}\gamma_5 s_b \, (\bar{u}_a \gamma^\nu C \bar{s}_b^T + \bar{u}_b \gamma^\nu C \bar{s}_a^T), \\[4pt]
		J_{6\mu} &= u_a^T C \gamma^\nu s_b \, (\bar{u}_a \sigma_{\mu\nu}\gamma_5 C \bar{s}_b^T - \bar{u}_b \sigma_{\mu\nu}\gamma_5 C \bar{s}_a^T) + u_a^T C \sigma_{\mu\nu}\gamma_5 s_b \, (\bar{u}_a \gamma^\nu C \bar{s}_b^T - \bar{u}_b \gamma^\nu C \bar{s}_a^T), \\[4pt]
		J_{7\mu} &= u_a^T C \gamma^\nu \gamma_5 s_b \, (\bar{u}_a \sigma_{\mu\nu} C \bar{s}_b^T + \bar{u}_b \sigma_{\mu\nu} C \bar{s}_a^T) + u_a^T C \sigma_{\mu\nu} s_b \, (\bar{u}_a \gamma^\nu \gamma_5 C \bar{s}_b^T + \bar{u}_b \gamma^\nu \gamma_5 C \bar{s}_a^T), \\[4pt]
		J_{8\mu} &= u_a^T C \gamma^\nu \gamma_5 s_b \, (\bar{u}_a \sigma_{\mu\nu} C \bar{s}_b^T - \bar{u}_b \sigma_{\mu\nu} C \bar{s}_a^T) + u_a^T C \sigma_{\mu\nu} s_b \, (\bar{u}_a \gamma^\nu \gamma_5 C \bar{s}_b^T - \bar{u}_b \gamma^\nu \gamma_5 C \bar{s}_a^T).
	\end{aligned}
\end{align}
All these operators possess the same overall quantum numbers as the $1^{++}$ tetraquark. When contracted in the two-point correlator, they generally induce both an axial-vector part related to the $1^{++}$ channel and a pseudoscalar part corresponding to a $0^{-+}$ component, as dictated by the decomposition shown in Sec.~II. In the following we will focus exclusively on the axial-vector piece, isolating the $1^{++}$ spectral function. 

With these currents at hand, we proceed to calculate the OPE of the correlators up to dimension-eight condensates and determine the corresponding spectral densities, which will serve as the input for the sum-rule analyses.

The operator product expansion (OPE) of the two-point correlation functions is performed according to the standard SVZ method~\cite{Shifman:1978bx}.  Truncating the expansion at dimension-8 and retaining the leading-order perturbative contribution, the LSR moment $M_1$ for the hidden-strange $us\bar{u}\bar{s}$ system can be written as
\begin{equation}
	\begin{aligned}
		M_{1}^{qs\bar{q}\bar{s}}(\tau, s_{0})
		= &\int_{4m_s^2}^{s_{0}} \rho_{1}^{qs\bar{q}\bar{s}}(s)\, e^{-\tau s}\,ds\\
		= &\int_{4m_s^2}^{s_0}\Biggl[
		\frac{1}{18432\pi^6} s^3 - \frac{m_s^2}{480\pi^6} s^2 
		+ \Bigl(-\frac{\langle \alpha_s GG\rangle}{4608\pi^5} 
		+ \frac{7m_s\langle\bar{q}q\rangle}{192\pi^4} 
		+ \frac{m_s\langle\bar{s}s\rangle}{64\pi^4} + \frac{3m_s^4}{256\pi^6}\Bigr) s \\
		& + \Bigl(-\frac{5\langle\bar{q}q\rangle\langle\bar{s}s\rangle}{18\pi^2} 
		+ \frac{5m_s\langle g_s\bar{q}\sigma G q\rangle}{96\pi^4} 
		- \frac{17m_s^2\langle \alpha_s GG\rangle}{9216\pi^5}-\frac{5m_s^3\langle \bar{q}q\rangle}{24\pi^4}-\frac{m_s^3\langle \bar{s}s\rangle}{12\pi^4}\Bigr)  \\
		& + \Bigl(-\frac{\langle\bar{q}q\rangle\langle g_s\bar{s}\sigma G s\rangle}{8\pi^2} 
		- \frac{\langle\bar{s}s\rangle\langle g_s\bar{q}\sigma G q\rangle}{8\pi^2} 
		+ \frac{m_s\langle \alpha_s GG\rangle\langle\bar{q}q\rangle}{128\pi^3} \\
		& + \frac{5m_s\langle \alpha_s GG\rangle\langle\bar{s}s\rangle}{384\pi^3} 
		+ \frac{m_s^2\langle\bar{q}q\rangle^2}{3\pi^2} 
		+ \frac{m_s^2\langle\bar{s}s\rangle^2}{24\pi^2} + \frac{3m_s^2\langle\bar{q}q\rangle\langle\bar{s}s\rangle}{4\pi^2}-\frac{3m_s^3\langle g_s\bar{q}\sigma Gq\rangle}{32\pi^4}\Bigr)\frac{1}{s}\Biggr] e^{-\tau s} ds
	\end{aligned}
\end{equation}
where we have added higher-order terms in $m_s$ compared to \cite{Chen:2015fwa,Chen:2013jra}. The LSR moments for other currents are listed in Appendix~\ref{Appendix:A}.
\section{Numerical Analysis for LSR and FESR}
\begin{figure}[htbp]
	\centering
	\subfigure[]{
		\includegraphics[scale=0.74]{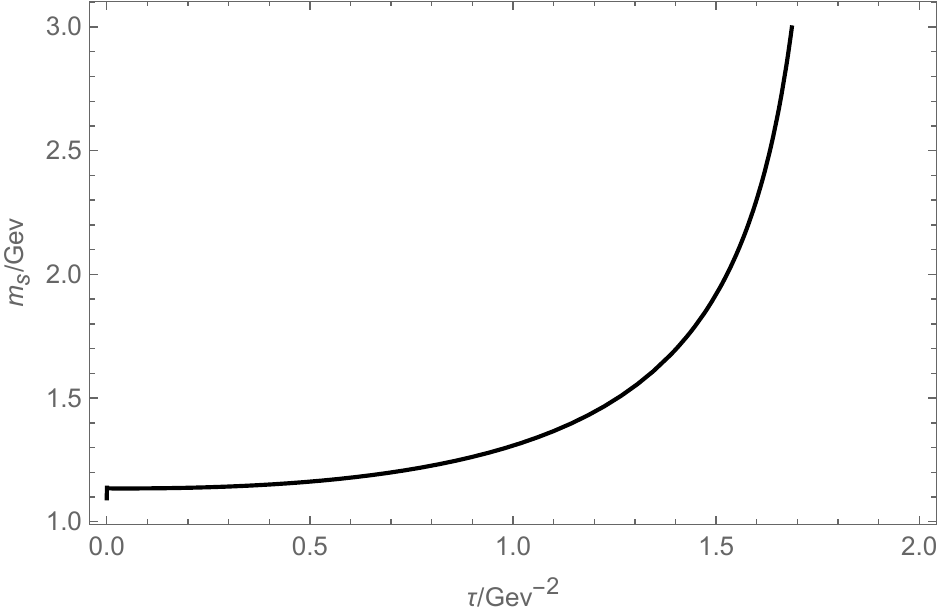}}
	\subfigure[]{
		\includegraphics[scale=0.7]{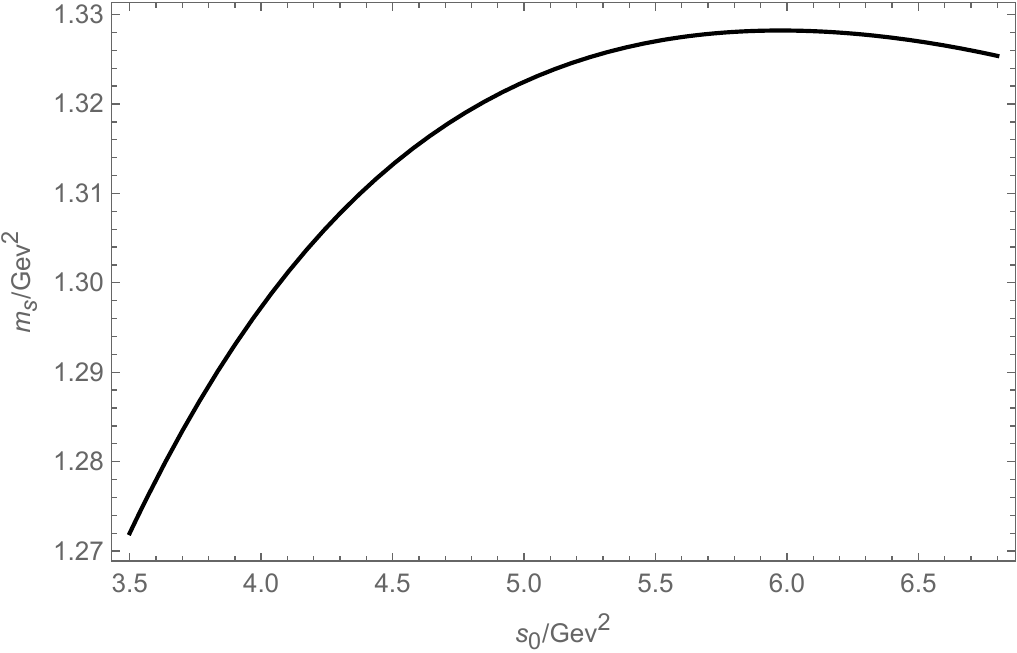}}
	\caption{\label{fig:LSR stability}(a) The $1^{++}$ four-quark masses versus
		$\tau$ obtained from the LSR for $J_{1\mu}$; (b) the same as (a) but for the masses versus $s_0$.}
\end{figure}

When performing the LSR analysis, it is essential to achieve stability in both the Borel parameter $\tau$ and the continuum threshold $s_0$ to obtain reliable predictions. In practice, we extract the hadron mass from the stationary points of the curves $m_H(\tau)$ and $m_H(s_0)$, which also allows a precise determination of the continuum threshold. Once LSR stability is attained, the convergence of the operator product expansion (OPE) at those stability points must be verified before the extracted mass values are included in the final average. For the $us\bar{u}\bar{s}$ interpolating currents, the LSR moment ratio derived from the current $J_1$ exhibits stable behavior, and its OPE series converges well: the highest included power corrections (dimension-8 condensates) amount to less than $20\%$ of the total contribution, justifying the truncation. Therefore, we adopt the result from $J_1$ for the final mass determination.
\begin{table}[htbp]
	\centering
	\caption{\label{tab:QCD parameters}QCD input parameters: $\rho$ represents the violation of factorization.}
	\begin{tabular}{ll}
		\hline
		Parameter & Value \\
		\hline
		$\langle \alpha_s G^2 \rangle$ & $(0.07 \pm 0.01)\ \mathrm{GeV}^4$ \\
		$\langle \bar{q} q \rangle$ & $(-0.24 \pm 0.01)^3\ \mathrm{GeV}^3$ \\
		$\langle \bar{s} s \rangle$ & $0.8\,\langle \bar{q} q \rangle$ \\
		$\langle \bar{q}\sigma G q \rangle$ & $m_0^2 \langle \bar{q} q \rangle$, $m_0^2 = (0.8 \pm 0.1)\ \mathrm{GeV}^2$ \\
		$\langle \bar{s} \sigma G s \rangle$ & $m_0^2 \langle \bar{s} s \rangle$ \\
		$m_s$ & $(95 \pm 5)\ \mathrm{MeV}$ \\
		$\rho$ & $2 \pm 1$ \\
		\hline
	\end{tabular}
\end{table}
In the numerical analysis, we adopt the QCD input parameters listed in Table~\ref{tab:QCD parameters}~\cite{Fu:2018ngx,Narison:2011xe,Narison:2014wqa}. Figure~\ref{fig:LSR stability} illustrates the LSR ratios for $J_1$: the upper panel displays the $\tau$-stability, while the lower panel shows the $s_0$-stability. It is evident that the LSR ratio for $J_1$ reaches both $\tau$ and $s_0$ stability when the dimension-8 condensate values are estimated using the vacuum saturation approximation. Moreover, we have verified that the mass curve remains stable even when the factorization assumption is violated by a factor of two, which provides an estimate of the theoretical uncertainty associated with such violation, complementing the central value. From Figure~\ref{fig:LSR stability}, the mass predictions at the $s_0$-stability extrema are read as
\begin{eqnarray}
	M_{\mathrm{LSR}} &=& 1.33(1.40)~\mathrm{GeV} \quad \text{at} \quad s_0 = 5.99(6.36)~\mathrm{GeV^2},
\end{eqnarray}
where the values in parentheses are obtained by allowing a factor-of-two violation of factorization for the dimension-8 condensates, and these will be used to estimate the errors.
\begin{figure}[htbp]
	\centering
	\includegraphics[scale=0.7]{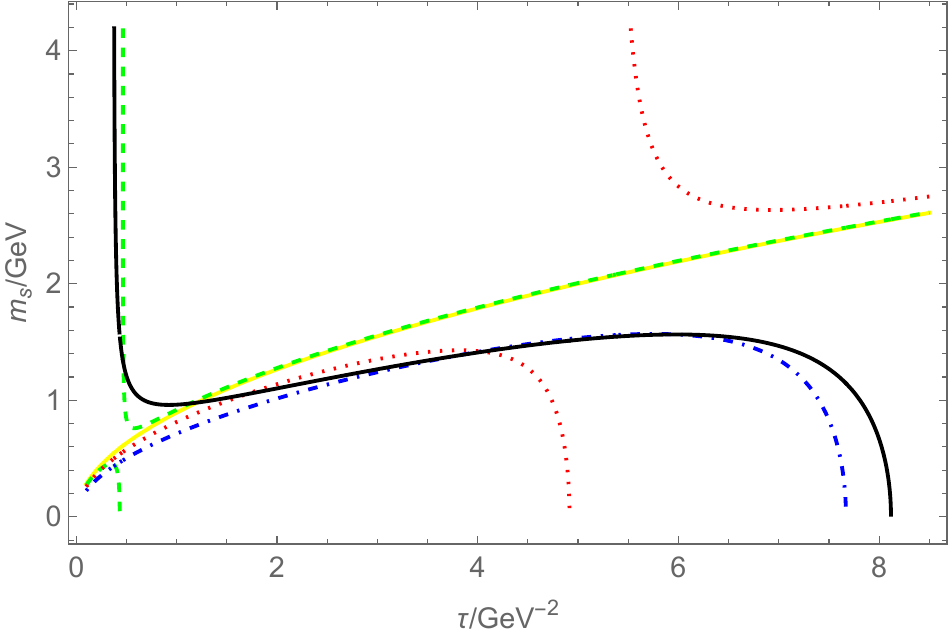}
	\caption{\label{fig:ususfesr} The mass of the $1^{++}$ $us\bar{u}\bar{s}$ tetraquark state as a function of $s_0$ in the FESR, obtained using $J_1$, with the OPE series contributions: perturbative term (yellow solid line), $d=2$ term (green dashed line), $d=4$ term (red dotted line), $d=6$ term (blue dash-dotted line), and $d=8$ term (black solid line).
	}
\end{figure}
Figure~\ref{fig:ususfesr} presents the FESR curves from OPE truncations at various orders. It is observed that, the corresponding FESR moment ratios rise gradually when only perturbative terms are taken into account. Once dimension-4 condensate contributions are included, the mass curves begin to exhibit inflection points (stability points), from which optimal mass values can be extracted. According to FIG.~\ref{fig:ususfesr}, the stability points of the FESR curves that include condensate terms up to dimension-8 (solid black lines) lie close to those derived from truncations at $d\leq6$ (blue dash-dotted lines). To clarify this observation, we note that using $J_1$ as interpolating currents, the inclusion of dimension-8 condensate contributions yields only a $1\%$ discrepancy in the extracted optimal masses, which is below $10\%$. This suggests that the OPE truncation for the FESR ratios corresponding to $J_1$ involve small uncertainties; therefore, in estimating the mass, we retain the results from $J_1$. The obtained values are as follows:
\begin{eqnarray}
	M_{FESR}&=&1.56(1.59)~{\rm GeV}~at~s_0=5.97(6.38)~{\rm GeV^2}.
\end{eqnarray}
By taking the arithmetic mean of the valid LSR and FESR results, while incorporating errors from QCD parameters and factorization violation, we obtain
\begin{eqnarray}
	M_{us\bar{u}\bar{s}}&=&1.45\pm0.11~{\rm GeV},
\end{eqnarray}
which agrees with the mass of the $a_1(1420)$ within the errors.

Upon combining the data from our analyses for different currents ($J_2$--$J_8$), we frequently obtain a mass estimate in the range of 1.5--1.8\,GeV. This outcome appears plausible. The existence of two nearby $a_1$ resonances, namely $a_1(1420)$ and $a_1(1640)$ or $a_1(1930)$, implies that if a given current couples to both states, yet the fitting procedure employs a single-pole approximation, the resulting pole mass will lie somewhere between the two actual masses.
\section{Gaussian Sum Rule for the $1^{++}$ Tetraquark}
In this section, we perform Gaussian sum rules for $J_1$ to estimate the mass for the first radial excited state of the $1^{++}$ hidden strange tetraquark\footnote{We did not perform the full GSR analysis for all eight currents ($J_2$-$J_8$) because the LSR/FESR results from those currents showed larger uncertainties and less stable behavior. }. To suppress the polynomial subtractions and enhance the resonance signal, we employ GSR defined as
\begin{equation}
	G(\hat{s},\tau) = \sqrt{\frac{\tau}{\pi}} \lim_{\substack{N,\Delta^2\to\infty\\ \tau=\Delta^2/(4N)}} \frac{(-\Delta^2)^N}{\Gamma(N)} \left(\frac{d}{d\Delta^2}\right)^N \left\{ \frac{\Pi(-\hat{s}-i\Delta) - \Pi(-\hat{s}+i\Delta)}{i\Delta} \right\}.
\end{equation}
After substituting the OPE result and performing the continuum subtraction, the subtracted GSR become
\begin{equation}
	G^{\mathrm{QCD}}(\hat{s},\tau,s_0) = \frac{1}{\sqrt{4\pi\tau}} \int_{4m_s^2}^{s_0} e^{-\frac{(\hat{s}-t)^2}{4\tau}} \, \rho^{\mathrm{QCD}}(t) \, dt,
\end{equation}
which, using the hadronic decomposition, also satisfies
\begin{equation}
	G^{\mathrm{QCD}}(\hat{s},\tau) = \frac{1}{\sqrt{4\pi\tau}} \int_{4m_s^2}^{\infty} e^{-\frac{(\hat{s}-t)^2}{4\tau}} \rho^{\mathrm{had}}(t) \, dt.
\end{equation}
To isolate the independent information, we define normalized GSR (NGSR):
\begin{equation}
	N^{\mathrm{QCD}}(\hat{s},\tau,s_0) = \frac{G^{\mathrm{QCD}}(\hat{s},\tau,s_0)}{\int_{-\infty}^{\infty} G^{\mathrm{QCD}}(\hat{s},\tau,s_0)\, d\hat{s}} = \frac{\frac{1}{\sqrt{4\pi\tau}}\int_{4m_s^2}^{\infty} e^{-\frac{(\hat{s}-t)^2}{4\tau}}\rho^{\mathrm{had}}(t)\,dt}{\int_{4m_s^2}^{\infty} \rho^{\mathrm{had}}(t)\,dt}.
\end{equation}
For the hadronic spectral function we adopt a two-resonance narrow-width model suitable for the \(1^{++}\) tetraquark channel:
\begin{equation}
	\rho^{\mathrm{had}}(t) = f_1^2 \delta(t-m_1^2) + f_2^2 \delta(t-m_2^2),
\end{equation}
where \(m_1\) corresponds to the light axial-vector tetraquark candidate (e.g., \(a_1(1420)\)) and \(m_2\) represents an additional, possibly heavier tetraquark state with the same $J^{PC}$ quantum number. The normalized couplings are
\begin{equation}
	r = \frac{f_1^2}{f_1^2+f_2^2}, \quad 1-r = \frac{f_2^2}{f_1^2+f_2^2}, \quad 0\leq r \leq 1.
\end{equation}
The corresponding hadronic NGSR becomes
\begin{equation}
	N^{\mathrm{had}}(\hat{s},\tau) = \frac{1}{\sqrt{4\pi\tau}} \left[ r\, e^{-\frac{(\hat{s}-m_1^2)^2}{4\tau}} + (1-r)\, e^{-\frac{(\hat{s}-m_2^2)^2}{4\tau}} \right].
\end{equation}

We fix \(m_1\) at the experimental mass of the best-known \(1^{++}\) light tetraquark candidate, e.g., \(a_1(1420)\) with \(m_1 = 1.420\) GeV, $s_0$ at the value determined by LSR and FESR, i.e. $5.8~\rm{GeV}^2$, and treat \(m_2\), \(r\) as fit parameters. The Gaussian width is chosen as \(\tau = 10\ \mathrm{GeV}^4\), sufficiently larger than the expected narrow widths of the states\footnote{The scale is chosen based on the compromise between suppressing duality violations and keeping $\tau$ sufficiently small\cite{Orlandini:2000nv}. we have verified that the results are stable against moderate variations of $\tau$ around this value. By choosing $ \tau = 8~\rm{GeV}^4$,
the mass would decrease by only 0.01 GeV.}. The fit minimizes
\begin{equation}
	\chi^2(r,m_2,s_0) = \sum_{\hat{s}_{\mathrm{min}}}^{\hat{s}_{\mathrm{max}}} \left[ N^{\mathrm{had}}(\hat{s},\tau) - N^{\mathrm{QCD}}(\hat{s},\tau,s_0) \right]^2
\end{equation}
over the range \(-10\ \mathrm{GeV}^2 \le \hat{s} \le 30\ \mathrm{GeV}^2\) with 161 equally spaced points. The optimized values obtained from the fit are
\begin{align}
	m_2 &= 1.86\pm 0.12\ \mathrm{GeV}, \\
	r &= 0.15\pm 0.03.
\end{align}
These results indicate that the coupling of the lighter state (\(m_1\)) to the tetraquark current is at most $15\%$ relative to the heavier one, suggesting that the \(a_1(1930)\) is a reasonable tetraquark candidate. In other words, the \(a_1(1420)\) may have a tetraquark component—otherwise the mass prediction would not agree—but that component is relatively small compared to the heavier state. This is not contradictory: it simply means that the \(a_1(1420)\) may have a mixed nature (e.g., a significant molecular or conventional $\bar{q}q$ component), while the \(a_1(1930)\) appears to be more “compact” and couples more strongly to the diquark–antidiquark current.
\section{Decay Patterns of the Heavier Tetraquark Candidate \(a_{1}(1930)\)}
In the previous section, the Gaussian sum rule analysis with a two-resonance model yielded a heavier state with mass \(m_{2} = 1.86 \pm 0.12\ \mathrm{GeV}\) and a coupling fraction \(r = 0.15 \pm 0.03\), indicating that the lighter state (identified with \(a_{1}(1420)\)) couples only weakly to the tetraquark current, while the heavier state dominates. This heavier mass is consistent with the \(a_{1}(1930)\) resonance listed by the Particle Data Group~\cite{ParticleDataGroup:2024cfk}. We now explore the possible decay modes of this heavier tetraquark candidate within the diquark–antidiquark picture.

The interpolating currents we constructed carry the quantum numbers \(J^{PC}=1^{++}\) and have a dominant \(us\bar{u}\bar{s}\) (hidden-strange) component. For a state with mass around \(1.86\ \mathrm{GeV}\), the phase space is significantly larger than that of the \(a_{1}(1420)\), opening up more hadronic decay channels. Using the Fierz transformation technique, the tetraquark current can be rearranged into colour-singlet meson pairs. For an isovector state with quark content \((us\bar{u}\bar{s}-ds\bar{d}\bar{s})/\sqrt{2}\), the possible two-meson intermediate states fall into two categories:
\begin{itemize}
	\item Channels with hidden strangeness (containing an \(\bar{s}s\) pair), such as \(K^{*}\bar{K}\) and \(f_{0}(980)\pi\), which are expected to be important due to the intrinsic strange component.
	\item Channels without hidden strangeness (light meson pairs), such as \(\rho\pi\), which may arise from the \(\bar{q}q\) components after Fierz transformation.
\end{itemize}

At \(m \approx 1.86\ \mathrm{GeV}\), the following two-body decays are kinematically allowed:
\begin{itemize}
	\item \textbf{\(a_{1}(1930) \to K^{*}(892)K\)}: The \(S\)-wave decay into a vector and a pseudoscalar meson is a natural channel for a \(1^{++}\) state. Both \(K^{*}\) and \(K\) contain one strange quark, conserving hidden strangeness. This channel is open (\(m_{K^{*}}+m_{K} \approx 1.39\ \mathrm{GeV}\)) and expected to be dominant.
	\item \textbf{\(a_{1}(1930) \to K_{1}(1270)K\)}: Decay into an axial-vector strange meson \(K_{1}(1270)\) (\(m \approx 1.27\ \mathrm{GeV}\)) plus a kaon is kinematically allowed and leaves ample phase space.
	\item \textbf{\(a_{1}(1930) \to K_{1}(1400)K\) and \(K^{*}(1410)K\)}: These channels involve heavier strange resonances (\(m_{K_{1}(1400)} \approx 1.40\ \mathrm{GeV}\), \(m_{K^{*}(1410)} \approx 1.41\ \mathrm{GeV}\)). Their thresholds are around \(1.89\)–\(1.90\ \mathrm{GeV}\), which lie slightly above the central value of our mass determination but could become accessible within the upper error range (\(m_{2} \lesssim 1.98\ \mathrm{GeV}\)). Therefore, these modes may contribute only marginally, depending on the actual mass of the state.
	\item \textbf{\(a_{1}(1930) \to \rho\pi\)}: This is a typical decay mode of light axial-vector mesons. Although it does not conserve hidden strangeness, it can arise from the \(\bar{q}q\) component after Fierz transformation, possibly with a reduced coupling. The \(\rho\pi\) channel has been observed in the decay of \(a_{1}(1260)\) and may also contribute for \(a_{1}(1930)\) due to the large phase space.
	\item \textbf{\(a_{1}(1930) \to f_{0}(980)\pi\)}: This is \(P\)-wave decays. The \(f_{0}(980)\) is scalar meson with a large \(\bar{s}s\) component, making this channel naturally coupled to the hidden-strange tetraquark. The \(f_{0}(980)\pi\) mode was crucial for the discovery of \(a_{1}(1420)\); for the heavier \(a_{1}(1930)\) the phase space is much larger, and this mode is expected to be prominent.
\end{itemize}

We note that the decay \(a_{1}(1930) \to K^{*}(1680)K\) is not kinematically allowed, as the threshold \(m_{K^{*}(1680)}+m_{K} \approx 2.21\ \mathrm{GeV}\) exceeds the mass of the candidate. Thus, such a channel should be excluded from the list of possible final states.

A quantitative estimate of the relative branching ratios requires detailed knowledge of the decay constants and form factors, which is beyond the scope of the present sum-rule analysis. Nevertheless, we can make a few qualitative observations:
\begin{enumerate}
	\item The \(S\)-wave \(K^{*}K\) and \(K\bar{K}^{*}\) channels are expected to be dominant because they require no orbital angular momentum and directly reflect the hidden-strange nature of the tetraquark.
	\item The \(P\)-wave decays into scalar mesons \((f_{0}(980)\pi)\) are also favoured by the presence of an \(\bar{s}s\) component in the final-state scalars.
	\item Decays into \(\rho\pi\) (light non-strange channels) may be suppressed relative to strange channels, but could still be sizeable due to the large phase space and mixing effects.
\end{enumerate}

Experimentally, the \(a_{1}(1930)\) has been observed in several production processes, but its decay pattern is not yet fully established. Future high-statistics experiments at BESIII, GlueX, and the upgraded COMPASS could search for the \(a_{1}(1930)\) in the \(K^{*}K\) and \(f_{0}(980)\pi\) final states. A comparison of the measured branching ratios with the expectations from the tetraquark picture would provide a crucial test of the nature of this state.

In summary, we propose that the heavier state extracted from the Gaussian sum rule, with mass around \(1.86\ \mathrm{GeV}\), be identified with the \(a_{1}(1930)\) resonance. Its dominant decay modes are predicted to be \(K^{*}K\) (\(S\)-wave) and \(f_{0}(980)\pi\) (\(P\)-wave), with possible sizeable contributions from \(K_{1}K\) and \(\rho\pi\) channels, while modes involving \(K_{1}(1400)\) or \(K^{*}(1410)\) may only be relevant if the mass is near the upper end of the uncertainty. Experimental searches for these decay modes are encouraged to verify the tetraquark interpretation.
\section{Summary and Conclusions}

In this work we have revisited the light \(1^{++}\) tetraquark sector using QCD sum rules, focusing on a complete set of derivative-free diquark-antidiquark interpolating currents. By employing both LSR and FESR, we have extracted the mass of the hidden-strange \(us\bar{u}\bar{s}\) tetraquark state. 

Our main results are summarized as follows:
\begin{itemize}
	\item For the hidden-strange channel, the combined LSR and FESR analysis gives 
	\[
	M_{us\bar{u}\bar{s}} = 1.45 \pm 0.11\ \mathrm{GeV},
	\]
	which is in good agreement with the experimental mass of the \(a_1(1420)\) resonance, supporting a possible tetraquark interpretation of this state.
	\item The GSR analysis, assuming a two-resonance narrow-width model, reveals a heavier state with mass 
	\[
	m_2 = 1.86 \pm 0.12\ \mathrm{GeV},
	\]
	and a relative coupling \(r = f_1^2/(f_1^2+f_2^2) = 0.15 \pm 0.03\). This indicates that the lighter \(a_1(1420)\) couples only weakly to the tetraquark current, while the heavier state, possibly identifiable with the \(a_1(1930)\), dominates the spectral function. Thus the \(a_1(1930)\) emerges as a promising candidate for a compact \(1^{++}\) tetraquark.
\end{itemize}

Regarding dominant decay modes, the \(us\bar{u}\bar{s}\) tetraquark is expected to decay predominantly into hidden-strange channels, such as \(K^*\bar{K}\) or \(K\bar{K}^*\), as well as into \(f_{0}(980)\pi\). Experimental searches for these decay patterns in high-statistics experiments (e.g., BESIII, LHCb) could further discriminate between tetraquark and conventional interpretations.

In conclusion, our QCD sum-rule analysis, based on derivative-free currents, provides strong evidence for the existence of light \(1^{++}\) tetraquarks and supports the assignment of the \(a_1(1420)\) and \(a_1(1930)\) as diquark-antidiquark states. Future lattice QCD simulations and experimental measurements of the predicted decay channels will be crucial to confirm this scenario.
\section*{Acknowledgements} 
This work is supported from the National Natural Science Foundation of China under Grant 12305104, 12665022, 12575153 and 12665009, and the Education Department of Hunan Province under Grant No. 24B0503.
\begin{appendix}
\section{LSR moments for $J_2$ to $J_8$}\label{Appendix:A}
In this appendix, we give LSR moments for $J_2$ to $J_8$ as follows
\begin{equation}
	\begin{aligned}
		M_{2}^{qs\bar{q}\bar{s}}(\tau, s_{0})
		= &\int_{4m_s^2}^{s_{0}} \rho_{2}^{qs\bar{q}\bar{s}}(s)\, e^{-\tau s}\,ds\\
		= &\int_{4m_s^2}^{s_0}\Biggl[
		\frac{1}{36864\pi^6} s^3 - \frac{m_s^2}{960\pi^6} s^2 
		+ \Bigl(\frac{\langle \alpha_s GG\rangle}{4608\pi^5} 
		+ \frac{7m_s\langle\bar{q}q\rangle}{384\pi^4} 
		+ \frac{m_s\langle\bar{s}s\rangle}{128\pi^4} + \frac{3m_s^4}{512\pi^6}\Bigr) s \\
		& + \Bigl(-\frac{5\langle\bar{q}q\rangle\langle\bar{s}s\rangle}{36\pi^2} 
		+ \frac{5m_s\langle g_s\bar{q}\sigma G q\rangle}{192\pi^4} 
		- \frac{13m_s^2\langle \alpha_s GG\rangle}{9216\pi^5}-\frac{5m_s^3\langle \bar{q}q\rangle}{48\pi^4}-\frac{m_s^3\langle \bar{s}s\rangle}{24\pi^4}\Bigr)  \\
		& + \Bigl(-\frac{\langle\bar{q}q\rangle\langle g_s\bar{s}\sigma G s\rangle}{16\pi^2} 
		- \frac{\langle\bar{s}s\rangle\langle g_s\bar{q}\sigma G q\rangle}{16\pi^2} 
		+ \frac{m_s\langle \alpha_s GG\rangle\langle\bar{q}q\rangle}{128\pi^3} \\
		& + \frac{m_s\langle \alpha_s GG\rangle\langle\bar{s}s\rangle}{384\pi^3} 
		+ \frac{m_s^2\langle\bar{q}q\rangle^2}{6\pi^2} 
		+ \frac{m_s^2\langle\bar{s}s\rangle^2}{48\pi^2}+ \frac{3m_s^2\langle\bar{q}q\rangle\langle\bar{s}s\rangle}{8\pi^2}-\frac{3m_s^3\langle g_s\bar{q}\sigma Gq\rangle}{64\pi^4}\Bigr)\frac{1}{s}\Biggr] e^{-\tau s} ds
	\end{aligned}
\end{equation}
\begin{equation}
	\begin{aligned}
		M_{3}^{qs\bar{q}\bar{s}}(\tau, s_{0})
		= &\int_{4m_s^2}^{s_{0}} \rho_{3}^{qs\bar{q}\bar{s}}(s)\, e^{-\tau s}\,ds\\
		= &\int_{4m_s^2}^{s_0}\Biggl[
		\frac{1}{18432\pi^6} s^3 - \frac{m_s^2}{480\pi^6} s^2 
		+ \Bigl(-\frac{\langle \alpha_s GG\rangle}{4608\pi^5} 
		- \frac{7m_s\langle\bar{q}q\rangle}{192\pi^4} 
		+ \frac{m_s\langle\bar{s}s\rangle}{64\pi^4} + \frac{3m_s^4}{256\pi^6}\Bigr) s \\
		& + \Bigl(\frac{5\langle\bar{q}q\rangle\langle\bar{s}s\rangle}{18\pi^2} 
		- \frac{5m_s\langle g_s\bar{q}\sigma G q\rangle}{96\pi^4} 
		- \frac{17m_s^2\langle \alpha_s GG\rangle}{9216\pi^5}+\frac{5m_s^3\langle \bar{q}q\rangle}{24\pi^4}-\frac{m_s^3\langle \bar{s}s\rangle}{12\pi^4}\Bigr)  \\
		& + \Bigl(\frac{\langle\bar{q}q\rangle\langle g_s\bar{s}\sigma G s\rangle}{8\pi^2} 
		+ \frac{\langle\bar{s}s\rangle\langle g_s\bar{q}\sigma G q\rangle}{8\pi^2} 
		- \frac{m_s\langle \alpha_s GG\rangle\langle\bar{q}q\rangle}{128\pi^3} \\
		& + \frac{5m_s\langle \alpha_s GG\rangle\langle\bar{s}s\rangle}{384\pi^3} 
		+ \frac{m_s^2\langle\bar{q}q\rangle^2}{3\pi^2} 
		+ \frac{m_s^2\langle\bar{s}s\rangle^2}{24\pi^2}- \frac{3m_s^2\langle\bar{q}q\rangle\langle\bar{s}s\rangle}{4\pi^2}+\frac{3m_s^3\langle g_s\bar{q}\sigma Gq\rangle}{32\pi^4}\Bigr)\frac{1}{s}\Biggr] e^{-\tau s} ds
	\end{aligned}
\end{equation}
\begin{equation}
	\begin{aligned}
		M_{4}^{qs\bar{q}\bar{s}}(\tau, s_{0})
		= &\int_{4m_s^2}^{s_{0}} \rho_{4}^{qs\bar{q}\bar{s}}(s)\, e^{-\tau s}\,ds\\
		= &\int_{4m_s^2}^{s_0}\Biggl[
		\frac{1}{36864\pi^6} s^3 - \frac{m_s^2}{960\pi^6} s^2 
		+ \Bigl(\frac{\langle \alpha_s GG\rangle}{4608\pi^5} 
		- \frac{7m_s\langle\bar{q}q\rangle}{384\pi^4} 
		+ \frac{m_s\langle\bar{s}s\rangle}{128\pi^4} + \frac{3m_s^4}{512\pi^6}\Bigr) s \\
		& + \Bigl(\frac{5\langle\bar{q}q\rangle\langle\bar{s}s\rangle}{36\pi^2} 
		- \frac{5m_s\langle g_s\bar{q}\sigma G q\rangle}{192\pi^4} 
		- \frac{13m_s^2\langle \alpha_s GG\rangle}{9216\pi^5}+\frac{5m_s^3\langle \bar{q}q\rangle}{48\pi^4}-\frac{m_s^3\langle \bar{s}s\rangle}{24\pi^4}\Bigr)  \\
		& + \Bigl(\frac{\langle\bar{q}q\rangle\langle g_s\bar{s}\sigma G s\rangle}{16\pi^2} 
		+ \frac{\langle\bar{s}s\rangle\langle g_s\bar{q}\sigma G q\rangle}{16\pi^2} 
		- \frac{m_s\langle \alpha_s GG\rangle\langle\bar{q}q\rangle}{128\pi^3} \\
		& + \frac{m_s\langle \alpha_s GG\rangle\langle\bar{s}s\rangle}{384\pi^3} 
		+ \frac{m_s^2\langle\bar{q}q\rangle^2}{6\pi^2} 
		+ \frac{m_s^2\langle\bar{s}s\rangle^2}{48\pi^2}- \frac{3m_s^2\langle\bar{q}q\rangle\langle\bar{s}s\rangle}{8\pi^2}+\frac{3m_s^3\langle g_s\bar{q}\sigma Gq\rangle}{64\pi^4}\Bigr)\frac{1}{s}\Biggr] e^{-\tau s} ds
	\end{aligned}
\end{equation}
\begin{equation}
	\begin{aligned}
		M_{5}^{qs\bar{q}\bar{s}}(\tau, s_{0})
		= &\int_{4m_s^2}^{s_{0}} \rho_{5}^{qs\bar{q}\bar{s}}(s)\, e^{-\tau s}\,ds\\
		= &\int_{4m_s^2}^{s_0}\Biggl[
		\frac{1}{6144\pi^6} s^3 - \frac{11m_s^2}{1280\pi^6} s^2 
		+ \Bigl(\frac{11\langle \alpha_s GG\rangle}{4608\pi^5} 
		+\frac{7m_s\langle\bar{q}q\rangle}{192\pi^4} 
		+ \frac{23m_s\langle\bar{s}s\rangle}{192\pi^4} + \frac{9m_s^4}{256\pi^6}\Bigr) s \\
		& + \Bigl(-\frac{5\langle\bar{q}q\rangle\langle\bar{s}s\rangle}{18\pi^2} 
		+ \frac{5m_s\langle g_s\bar{q}\sigma G q\rangle}{96\pi^4} 
		- \frac{163m_s^2\langle \alpha_s GG\rangle}{9216\pi^5}-\frac{5m_s^3\langle \bar{q}q\rangle}{24\pi^4}-\frac{m_s^3\langle \bar{s}s\rangle}{4\pi^4}+ \\
		& + \frac{5m_s\langle g_s\bar{s}\sigma G s\rangle}{48\pi^4}
		-\frac{5\langle\bar{q}q\rangle^2}{18\pi^2}-\frac{5\langle\bar{s}s\rangle^2}{18\pi^2} \Bigr)  \\
		& + \Bigl(-\frac{\langle\bar{q}q\rangle\langle g_s\bar{s}\sigma G s\rangle}{8\pi^2} 
		- \frac{\langle\bar{s}s\rangle\langle g_s\bar{q}\sigma G q\rangle}{8\pi^2} 
		+ \frac{3m_s\langle \alpha_s GG\rangle\langle\bar{q}q\rangle}{128\pi^3} \\
		& + \frac{5m_s\langle \alpha_s GG\rangle\langle\bar{s}s\rangle}{128\pi^3} 
		+ \frac{2m_s^2\langle\bar{q}q\rangle^2}{\pi^2} 
		+ \frac{m_s^2\langle\bar{s}s\rangle^2}{8\pi^2}+ \frac{3m_s^2\langle\bar{q}q\rangle\langle\bar{s}s\rangle}{4\pi^2}-\frac{3m_s^3\langle g_s\bar{q}\sigma Gq\rangle}{32\pi^4}\\
		& - \frac{\langle\bar{q}q\rangle\langle g_s\bar{q}\sigma G q\rangle}{4\pi^2}-\frac{\langle\bar{s}s\rangle\langle g_s\bar{s}\sigma G s\rangle}{4\pi^2} \Bigr)\frac{1}{s}\Biggr] e^{-\tau s} ds
	\end{aligned}
\end{equation}
\begin{equation}
	\begin{aligned}
		M_{6}^{qs\bar{q}\bar{s}}(\tau, s_{0})
		= &\int_{4m_s^2}^{s_{0}} \rho_{6}^{qs\bar{q}\bar{s}}(s)\, e^{-\tau s}\,ds\\
		= &\int_{4m_s^2}^{s_0}\Biggl[
		\frac{1}{12288\pi^6} s^3 - \frac{11m_s^2}{2560\pi^6} s^2 
		+ \Bigl(\frac{\langle \alpha_s GG\rangle}{4608\pi^5} 
		+\frac{7m_s\langle\bar{q}q\rangle}{384\pi^4} 
		+ \frac{23m_s\langle\bar{s}s\rangle}{384\pi^4} + \frac{9m_s^4}{512\pi^6}\Bigr) s \\
		& + \Bigl(-\frac{5\langle\bar{q}q\rangle\langle\bar{s}s\rangle}{36\pi^2} 
		+ \frac{5m_s\langle g_s\bar{q}\sigma G q\rangle}{192\pi^4} 
		- \frac{23m_s^2\langle \alpha_s GG\rangle}{9216\pi^5}-\frac{5m_s^3\langle \bar{q}q\rangle}{48\pi^4}-\frac{m_s^3\langle \bar{s}s\rangle}{8\pi^4}+ \\
		& + \frac{5m_s\langle g_s\bar{s}\sigma G s\rangle}{96\pi^4}
		-\frac{5\langle\bar{q}q\rangle^2}{36\pi^2}-\frac{5\langle\bar{s}s\rangle^2}{36\pi^2} \Bigr)  \\
		& + \Bigl(-\frac{\langle\bar{q}q\rangle\langle g_s\bar{s}\sigma G s\rangle}{16\pi^2} 
		- \frac{\langle\bar{s}s\rangle\langle g_s\bar{q}\sigma G q\rangle}{16\pi^2} 
		+ \frac{3m_s\langle \alpha_s GG\rangle\langle\bar{q}q\rangle}{128\pi^3} \\
		& + \frac{m_s\langle \alpha_s GG\rangle\langle\bar{s}s\rangle}{128\pi^3} 
		+ \frac{m_s^2\langle\bar{q}q\rangle^2}{\pi^2} 
		+ \frac{m_s^2\langle\bar{s}s\rangle^2}{16\pi^2}+ \frac{3m_s^2\langle\bar{q}q\rangle\langle\bar{s}s\rangle}{8\pi^2}-\frac{3m_s^3\langle g_s\bar{q}\sigma Gq\rangle}{64\pi^4}\\
		& - \frac{\langle\bar{q}q\rangle\langle g_s\bar{q}\sigma G q\rangle}{8\pi^2}-\frac{\langle\bar{s}s\rangle\langle g_s\bar{s}\sigma G s\rangle}{8\pi^2} \Bigr)\frac{1}{s}\Biggr] e^{-\tau s} ds
	\end{aligned}
\end{equation}
\begin{equation}
	\begin{aligned}
		M_{7}^{qs\bar{q}\bar{s}}(\tau, s_{0})
		= &\int_{4m_s^2}^{s_{0}} \rho_{7}^{qs\bar{q}\bar{s}}(s)\, e^{-\tau s}\,ds\\
		= &\int_{4m_s^2}^{s_0}\Biggl[
		\frac{1}{6144\pi^6} s^3 - \frac{11m_s^2}{1280\pi^6} s^2 
		+ \Bigl(\frac{11\langle \alpha_s GG\rangle}{4608\pi^5} 
		-\frac{7m_s\langle\bar{q}q\rangle}{192\pi^4} 
		+ \frac{23m_s\langle\bar{s}s\rangle}{192\pi^4} + \frac{9m_s^4}{256\pi^6}\Bigr) s \\
		& + \Bigl(\frac{5\langle\bar{q}q\rangle\langle\bar{s}s\rangle}{18\pi^2} 
		- \frac{5m_s\langle g_s\bar{q}\sigma G q\rangle}{96\pi^4} 
		- \frac{163m_s^2\langle \alpha_s GG\rangle}{9216\pi^5}+\frac{5m_s^3\langle \bar{q}q\rangle}{24\pi^4}-\frac{m_s^3\langle \bar{s}s\rangle}{4\pi^4}+ \\
		& + \frac{5m_s\langle g_s\bar{s}\sigma G s\rangle}{48\pi^4}
		-\frac{5\langle\bar{q}q\rangle^2}{18\pi^2}-\frac{5\langle\bar{s}s\rangle^2}{18\pi^2} \Bigr)  \\
		& + \Bigl(\frac{\langle\bar{q}q\rangle\langle g_s\bar{s}\sigma G s\rangle}{8\pi^2} 
		+ \frac{\langle\bar{s}s\rangle\langle g_s\bar{q}\sigma G q\rangle}{8\pi^2} 
		- \frac{3m_s\langle \alpha_s GG\rangle\langle\bar{q}q\rangle}{128\pi^3} \\
		& + \frac{5m_s\langle \alpha_s GG\rangle\langle\bar{s}s\rangle}{128\pi^3} 
		+ \frac{2m_s^2\langle\bar{q}q\rangle^2}{\pi^2} 
		+ \frac{m_s^2\langle\bar{s}s\rangle^2}{8\pi^2}- \frac{3m_s^2\langle\bar{q}q\rangle\langle\bar{s}s\rangle}{4\pi^2}+\frac{3m_s^3\langle g_s\bar{q}\sigma Gq\rangle}{32\pi^4}\\
		& - \frac{\langle\bar{q}q\rangle\langle g_s\bar{q}\sigma G q\rangle}{4\pi^2}-\frac{\langle\bar{s}s\rangle\langle g_s\bar{s}\sigma G s\rangle}{4\pi^2} \Bigr)\frac{1}{s}\Biggr] e^{-\tau s} ds
	\end{aligned}
\end{equation}
\begin{equation}
	\begin{aligned}
		M_{8}^{qs\bar{q}\bar{s}}(\tau, s_{0})
		= &\int_{4m_s^2}^{s_{0}} \rho_{8}^{qs\bar{q}\bar{s}}(s)\, e^{-\tau s}\,ds\\
		= &\int_{4m_s^2}^{s_0}\Biggl[
		\frac{1}{12288\pi^6} s^3 - \frac{11m_s^2}{2560\pi^6} s^2 
		+ \Bigl(\frac{\langle \alpha_s GG\rangle}{4608\pi^5} 
		-\frac{7m_s\langle\bar{q}q\rangle}{384\pi^4} 
		+ \frac{23m_s\langle\bar{s}s\rangle}{384\pi^4} + \frac{9m_s^4}{512\pi^6}\Bigr) s \\
		& + \Bigl(\frac{5\langle\bar{q}q\rangle\langle\bar{s}s\rangle}{36\pi^2} 
		- \frac{5m_s\langle g_s\bar{q}\sigma G q\rangle}{192\pi^4} 
		- \frac{23m_s^2\langle \alpha_s GG\rangle}{9216\pi^5}+\frac{5m_s^3\langle \bar{q}q\rangle}{48\pi^4}-\frac{m_s^3\langle \bar{s}s\rangle}{8\pi^4}+ \\
		& + \frac{5m_s\langle g_s\bar{s}\sigma G s\rangle}{96\pi^4}
		-\frac{5\langle\bar{q}q\rangle^2}{36\pi^2}-\frac{5\langle\bar{s}s\rangle^2}{36\pi^2} \Bigr)  \\
		& + \Bigl(\frac{\langle\bar{q}q\rangle\langle g_s\bar{s}\sigma G s\rangle}{16\pi^2} 
		+ \frac{\langle\bar{s}s\rangle\langle g_s\bar{q}\sigma G q\rangle}{16\pi^2} 
		- \frac{3m_s\langle \alpha_s GG\rangle\langle\bar{q}q\rangle}{128\pi^3} \\
		& + \frac{m_s\langle \alpha_s GG\rangle\langle\bar{s}s\rangle}{128\pi^3} 
		+ \frac{m_s^2\langle\bar{q}q\rangle^2}{\pi^2} 
		+ \frac{m_s^2\langle\bar{s}s\rangle^2}{16\pi^2}- \frac{3m_s^2\langle\bar{q}q\rangle\langle\bar{s}s\rangle}{8\pi^2}+\frac{3m_s^3\langle g_s\bar{q}\sigma Gq\rangle}{64\pi^4}\\
		& - \frac{\langle\bar{q}q\rangle\langle g_s\bar{q}\sigma G q\rangle}{8\pi^2}-\frac{\langle\bar{s}s\rangle\langle g_s\bar{s}\sigma G s\rangle}{8\pi^2} \Bigr)\frac{1}{s}\Biggr] e^{-\tau s} ds
	\end{aligned}
\end{equation}
\end{appendix}
\bibliographystyle{plain}
\bibliography{ref}

@article{Chen:2015fwa,
    author = "Chen, Hua-Xing and Cui, Er-Liang and Chen, Wei and Steele, T. G. and Liu, Xiang and Zhu, Shi-Lin",
    title = "{$a_1(1420)$ resonance as a tetraquark state and its isospin partner}",
    eprint = "1503.02597",
    archivePrefix = "arXiv",
    primaryClass = "hep-ph",
    doi = "10.1103/PhysRevD.91.094022",
    journal = "Phys. Rev. D",
    volume = "91",
    pages = "094022",
    year = "2015"
}

@article{COMPASS:2015kdx,
    author = "Adolph, C. and others",
    collaboration = "COMPASS",
    title = "{Observation of a New Narrow Axial-Vector Meson $a_1$(1420)}",
    eprint = "1501.05732",
    archivePrefix = "arXiv",
    primaryClass = "hep-ex",
    reportNumber = "CERN-PH-EP-2015-015",
    doi = "10.1103/PhysRevLett.115.082001",
    journal = "Phys. Rev. Lett.",
    volume = "115",
    number = "8",
    pages = "082001",
    year = "2015"
}

@article{Shifman:1978bx,
    author = "Shifman, Mikhail A. and Vainshtein, A. I. and Zakharov, Valentin I.",
    title = "{QCD and Resonance Physics. Theoretical Foundations}",
    reportNumber = "ITEP-73-1978, ITEP-80-1978",
    doi = "10.1016/0550-3213(79)90022-1",
    journal = "Nucl. Phys. B",
    volume = "147",
    pages = "385--447",
    year = "1979"
}

@article{Reinders:1984sr,
    author = "Reinders, L. J. and Rubinstein, H. and Yazaki, S.",
    title = "{Hadron Properties from QCD Sum Rules}",
    reportNumber = "CERN-TH-4079-84",
    doi = "10.1016/0370-1573(85)90065-1",
    journal = "Phys. Rept.",
    volume = "127",
    pages = "1",
    year = "1985"
}

@article{Colangelo:2000dp,
    author = "Colangelo, Pietro and Khodjamirian, Alexander",
    editor = "Shifman, M. and Ioffe, Boris",
    title = "{QCD sum rules, a modern perspective}",
    eprint = "hep-ph/0010175",
    archivePrefix = "arXiv",
    reportNumber = "CERN-TH-2000-296, BARI-TH-2000-394",
    doi = "10.1142/9789812810458_0033",
    pages = "1495--1576",
    month = "10",
    year = "2000"
}

@article{Chen:2013jra,
    author = "Chen, Hua-Xing",
    title = "{Chiral Structure of Vector and Axial-Vector Tetraquark Currents}",
    eprint = "1311.4992",
    archivePrefix = "arXiv",
    primaryClass = "hep-ph",
    doi = "10.1140/epjc/s10052-013-2628-y",
    journal = "Eur. Phys. J. C",
    volume = "73",
    pages = "2628",
    year = "2013"
}

@article{Huang:2016rro,
    author = "Huang, Zhuo-Ran and Chen, Wei and Steele, T. G. and Zhang, Zhu-Feng and Jin, Hong-Ying",
    title = "{Investigation of the light four-quark states with exotic $J^{PC}=0^{--}$}",
    eprint = "1610.02081",
    archivePrefix = "arXiv",
    primaryClass = "hep-ph",
    doi = "10.1103/PhysRevD.95.076017",
    journal = "Phys. Rev. D",
    volume = "95",
    number = "7",
    pages = "076017",
    year = "2017"
}

@article{Fu:2018ngx,
    author = "Fu, Yi-Chao and Huang, Zhuo-Ran and Zhang, Zhu-Feng and Chen, Wei",
    title = "{Exotic tetraquark states with $J^{PC}=0^{+-}$}",
    eprint = "1811.03333",
    archivePrefix = "arXiv",
    primaryClass = "hep-ph",
    doi = "10.1103/PhysRevD.99.014025",
    journal = "Phys. Rev. D",
    volume = "99",
    number = "1",
    pages = "014025",
    year = "2019"
}

@article{Sundu:2017xct,
    author = "Sundu, H. and Agaev, S. S. and Azizi, K.",
    title = "{New $\alpha_1$(1420) state: Structure, mass, and width}",
    eprint = "1711.05977",
    archivePrefix = "arXiv",
    primaryClass = "hep-ph",
    doi = "10.1103/PhysRevD.97.054001",
    journal = "Phys. Rev. D",
    volume = "97",
    number = "5",
    pages = "054001",
    year = "2018"
}

@article{Chen:2008qw,
    author = "Chen, Hua-Xing and Hosaka, Atsushi and Zhu, Shi-Lin",
    title = "{The I**G J**PC = 1- 1-+ Tetraquark States}",
    eprint = "0806.1998",
    archivePrefix = "arXiv",
    primaryClass = "hep-ph",
    doi = "10.1103/PhysRevD.78.054017",
    journal = "Phys. Rev. D",
    volume = "78",
    pages = "054017",
    year = "2008"
}

@article{ParticleDataGroup:2024cfk,
    author = "Navas, S. and others",
    collaboration = "Particle Data Group",
    title = "{Review of particle physics}",
    doi = "10.1103/PhysRevD.110.030001",
    journal = "Phys. Rev. D",
    volume = "110",
    number = "3",
    pages = "030001",
    year = "2024"
}

@article{Chen:2010ze,
    author = "Chen, Wei and Zhu, Shi-Lin",
    title = "{The Vector and Axial-Vector Charmonium-like States}",
    eprint = "1010.3397",
    archivePrefix = "arXiv",
    primaryClass = "hep-ph",
    doi = "10.1103/PhysRevD.83.034010",
    journal = "Phys. Rev. D",
    volume = "83",
    pages = "034010",
    year = "2011"
}

@article{Ho:2019org,
    author = "Ho, J. and Berg, R. and Steele, T. G. and Chen, W. and Harnett, D.",
    title = "{Is the $Y(2175)$ a Strangeonium Hybrid Meson?}",
    eprint = "1905.12779",
    archivePrefix = "arXiv",
    primaryClass = "hep-ph",
    doi = "10.1103/PhysRevD.100.034012",
    journal = "Phys. Rev. D",
    volume = "100",
    number = "3",
    pages = "034012",
    year = "2019"
}

@article{Ho:2018cat,
    author = "Ho, J. and Berg, R. and Chen, Wei and Harnett, D. and Steele, T. G.",
    title = "{Mass Calculations of Light Quarkonium, Exotic $J^{PC}=0^{+-}$ Hybrid Mesons from Gaussian Sum-Rules}",
    eprint = "1806.02465",
    archivePrefix = "arXiv",
    primaryClass = "hep-ph",
    doi = "10.1103/PhysRevD.98.096020",
    journal = "Phys. Rev. D",
    volume = "98",
    number = "9",
    pages = "096020",
    year = "2018"
}

@article{Narison:2011xe,
    author = "Narison, Stephan",
    title = "{Gluon Condensates and precise $\overline{m}_{c,b}$ from QCD-Moments and their ratios to Order $\alpha_s^3$ and {\ensuremath{<}} G$^4$ {\ensuremath{>}}}",
    eprint = "1105.2922",
    archivePrefix = "arXiv",
    primaryClass = "hep-ph",
    doi = "10.1016/j.physletb.2011.11.058",
    journal = "Phys. Lett. B",
    volume = "706",
    pages = "412--422",
    year = "2012"
}

@article{Narison:2014wqa,
    author = "Narison, Stephan",
    editor = "Narison, Stephan",
    title = "{Mini-review on QCD spectral sum rules}",
    eprint = "1409.8148",
    archivePrefix = "arXiv",
    primaryClass = "hep-ph",
    doi = "10.1016/j.nuclphysbps.2015.01.041",
    journal = "Nucl. Part. Phys. Proc.",
    volume = "258-259",
    pages = "189--194",
    year = "2015"
}

@article{Mikhasenko2015,
  author = {M. Mikhasenko and B. Ketzer and Andrey Sarantsev},
  title = {Nature of the \(a_1(1420)\)},
  journal = {Phys. Rev. D},
  volume = {91},
  pages = {094015},
  year = {2015},
  eprint = {1501.07023},
  archivePrefix = {arXiv},
  primaryClass = {hep-ph}
}

@article{Aceti2016,
  author = {F. Aceti and L. R. Dai and E. Oset},
  title = {\(a_1(1420)\) peak as the \(\pi f_0(980)\) decay mode of the \(a_1(1260)\)},
  journal = {Phys. Rev. D},
  volume = {94},
  pages = {096015},
  year = {2016},
  eprint = {1606.06893},
  archivePrefix = {arXiv},
  primaryClass = {hep-ph}
}

@article{Guo2020,
  author = {Feng-Kun Guo and Xiao-Hai Liu and Shuntaro Sakai},
  title = {Threshold cusps and triangle singularities in hadronic reactions},
  journal = {Prog. Part. Nucl. Phys.},
  volume = {112},
  pages = {103757},
  year = {2020},
  eprint = {1912.07030},
  archivePrefix = {arXiv},
  primaryClass = {hep-ph}
}

@article{Sakthivasan2024,
  author = {Ajay S. Sakthivasan and Maxim Mai and Akaki Rusetsky and Michael Doring},
  title = {Effects of final state interactions on Landau singularities},
  journal = {JHEP},
  volume = {10},
  pages = {246},
  year = {2024},
  eprint = {2407.17969},
  archivePrefix = {arXiv},
  primaryClass = {hep-ph}
}

@article{Sakthivasan:2026bph,
    author = {Sakthivasan, Ajay S. and Feng, Yuchuan and D{\"o}ring, Michael and Mai, Maxim},
    title = "{The $a_1(1420)$ in a Unitary Coupled-Channel Three-Body Approach}",
    eprint = "2606.24709",
    archivePrefix = "arXiv",
    primaryClass = "hep-ph",
    reportNumber = "JLAB-THY-26-4812",
    month = "6",
    year = "2026"
}

@article{Yan2025,
  author = {Mao-Jun Yan and Chun-Sheng An and Cheng-Rong Deng},
  title = {Reevaluating the \(a_1(1420)\) enhancement and its molecular partners in the low-lying axial-vector meson spectrum},
  year = {2025},
  eprint = {2512.02655},
  archivePrefix = {arXiv},
  primaryClass = {hep-ph}
}

@article{Orlandini:2000nv,
    author = "Orlandini, G. and Steele, Thomas G. and Harnett, D.",
    title = "{Gaussian sum rules and prediction of resonance properties}",
    eprint = "hep-ph/0007299",
    archivePrefix = "arXiv",
    doi = "10.1016/S0375-9474(00)00512-1",
    journal = "Nucl. Phys. A",
    volume = "686",
    pages = "261--289",
    year = "2001"
}
\end{document}